\documentclass[preprintnumbers,twocolumn,amsmath,amssymb,floats]{revtex4}
\usepackage{graphicx}
\begin{document}

\preprint{Presented at PLDS-3}

\title{Luttinger-liquid-like behavior in bulk crystals of 
the quasi-one-dimensional conductor NbSe$_3$}

\author{S.V.~Zaitsev-Zotov$^a$, M.S.H.~Go$^b$ E.~Slot$^b$, H.S.J. van der
Zant$^b$\\}
\address{$^a$Institute of
Radioengineering and Electronics of Russian Academy of Sciences, Mokhovaya 11,
103907 Moscow, Russia\\}
\address{$^b$DIMES and Department of Applied Sciences, Delft University of 
Technology, Lorentzweg 1, 2628 CJ Delft, The Netherlands\\}
\begin{abstract}
{CDW/Normal metal/CDW junctions and nanoconstrictions in crystals of the
quasi-one-dimensional conductor NbSe$_3$ are manufactured using a
focused-ion-beam. It is found that the low-temperature conduction of these
structures changes dramatically and loses the features of the
charge-density-wave transition. Instead, a dielectric phase is developed.
Up to 6-order power-law variations of the conduction as a function of both
temperature and electric field can be observed for this new phase.  The
transition from quasi-one-dimensional behavior to one-dimensional
behavior is associated with destruction of the three-dimensional order of the
charge-density waves by fluctuations. It results in a recovery of the Luttinger-liquid
properties of metallic chains, like it takes place in sliding Luttinger liquid
phase.}
\end{abstract}

\maketitle

\section{Introduction}

Transport properties of electron systems strongly depend on dimensionality.
In one dimension the Coulomb interaction between carriers modifies the ground
state and dramatically changes the transport properties.  In a one-dimensional
(1D) metal, instead of a Fermi liquid of electrons, a so-called Luttinger liquid
(LL) is developed~\cite{LLReview}.  Elementary excitations in a LL are collective
charge and spin boson modes, rather than fermion quasiparticles as in 3D metals.
The ground state of a LL is characterized by a power-law suppression of the 
tunneling density of states.  Impurities in a LL act as tunneling barriers. 
Therefore, the conduction $G$ of impure LLs is a power-law function of 
temperature, $G\propto T^\alpha$, and voltage, $G\propto V^{\beta-1}$~($I\propto
V^{\beta}$ in current-voltage curves)~\cite{KF}.
These dependencies are considered as a fingerprint of 1D behavior.

The experimental study of 1D effects in long 
nanowires remains a challenge. The main efforts are focused on developing a
technology to fabricate metal or semiconductor quantum wires with
transverse sizes on the order of 10 nm or less and attaching them to an 
external circuit.  At the present, both problems were successfully solved for a number of
physical systems including single-wall carbon nanotubes~\cite{carbon},
Mo$_6$O$_6$ molecular wires~\cite{Mo6O6}, Si whiskers~\cite{Si}, InSb nanowires
in asbestos matrices~\cite{InSb}, Bi nanowires in various matrixes (see e.g.
\cite{Bi_Al2O3} and references therein), {\it etc}.  Experimental observation of
the power-law-supporting LL theories has been reported for single-wall carbon
nanotubes~\cite{carbon,Sonin}, Mo$_6$O$_6$ nanowires~\cite{Mo6O6nano} and InSb
nanowires~\cite{InSb}.  In all cases the nanowire diameters do not exceed 10 
nm.

There is a wide class of materials suitable for experimental search and study of
1D effects: quasi-one dimensional (quasi-1D) conductors with a charge-density
wave (CDW) ground state~\cite{CDWReview}.  Such materials consist of metallic
chains connected in a crystal through the relatively weak Van der Waals
interaction.  At temperatures below the Peierls transition temperature, $T_p$,
the Peierls instability leads to the development of the CDW, which is a 3D-ordered
periodic modulation of the electron density accompanied by a periodic distortion
of the crystalline lattice. 

An example of these materials is NbSe$_3$.  It has a monoclinic unit cell with 
lattice parameters $a=10.01$~\AA, $b=3.48$~\AA, $c=15.63$~\AA, and
$\beta=109.5^o$~\cite{CDWReview}.  The conducting chains are along the $b$
direction.  NbSe$_3$ exhibits two Peierls transitions leading to the partial
dielectrization of the electron spectrum, at $T_{P1}=145$~K and $T_{P2}=59$~K. 
The remaining non-condensed electrons provide metallic conduction down to the 
lowest temperatures.

In some sense, the CDW state results from an instability of LLs with respect to 3D coupling at low enough temperatures.  It is also widely accepted that in
quasi-1D conductors LL exists at temperatures above $T_P$ (see {\it e.g.}
\cite{Voit}).  As $T_p$ decreases with the sample cross-section
area~\cite{BZZN}, one could expect the area of existence of LL to grow towards
the low-temperature region for sufficiently thin crystals.

Transitions from quasi-1D to 1D behavior with reduction of the transverse sample
sizes has recently been observed in the quasi-1D conductors NbSe$_3$ and
TaS$_3$~\cite{ZZPM}.  Here we present the results of our observation of even 
more dramatic changes in the conduction of quasi-1D conductors found in two 
novel (for quasi-1D conductors) types of structures:  CDW/normal-metal/CDW
(C/N/C) junctions and nanoconstrictions.  Surprisingly, 1D behavior has been
observed in samples with transverse sizes in the $\mu$m-range scale.

\section{Technology}

Structures were prepared at the Delft Institute for Microfabrication and
Submicron technology (DIMES).  We used a standard Fei Inc.  200xP focus-ion-beam
(FIB) system with a gallium ion source.  This FIB system was used for imaging,
etching and platinum deposition {\it in situ}.  In all cases the FIB was used at
30 kV.  When imaging, the lowest beam currents were used ($I=1$ - 4 pA) to
minimize possible damage of incoming gallium ions.

\begin{figure}
\includegraphics[width=8cm]{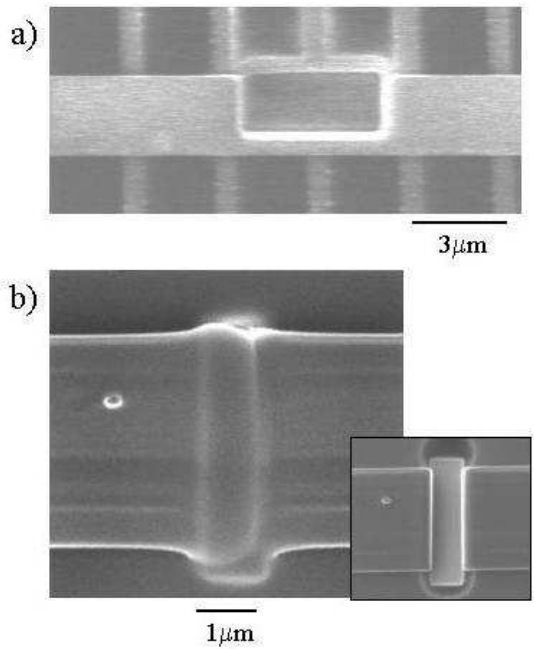}
\caption{a) FIB fabricated nanoconstriction in NbSe$_3$ crystal.
b) FIB fabrication of a C/N/C junction. Inset: a gap of 1~$\mu$m width is
etched into a crystal. Left: the gap is filled with platinum.}
\label{FIB}
\end{figure}

For the fabrication of the CDW devices, a long, narrow and flat 
quasi-1D crystal is placed on top of a sapphire substrate with preliminary 
deposited gold contacts. For nanoconstriction fabrication, the FIB is used
to remove part of a crystal between two gold terminals (Fig.~\ref{FIB}(a)).

C/N/C junctions were fabricated in two steps. In the first step the FIB is
used to make a gap in the crystal (right part of Fig.~\ref{FIB}(b)). In the
second step the gap is filled {\it in situ} with platinum by FIB deposition
(left part of Fig.~\ref{FIB}(b)).  In some cases additional platinum voltage
terminals were attached to the crystal using FIB platinum deposition.

\section{Results}

\subsection{Nanoconstrictions}

Damage during etching may have important consequences for the interpretation of
the data.  However, up to now we have not found any indication for serious
damage to our samples due to FIB milling.  For example, Shapiro step
measurements on samples etched with the FIB still show complete mode-locking
indicating their good quality.  In addition, for NbSe$_3$ the resistance ratios,
$RRR=R(300{\rm~K})/R(4.2{\rm~K})$, of etched sample parts are the same as the
ones of unetched parts.

\begin{figure}
\vskip -1.5cm
\includegraphics[width=10cm]{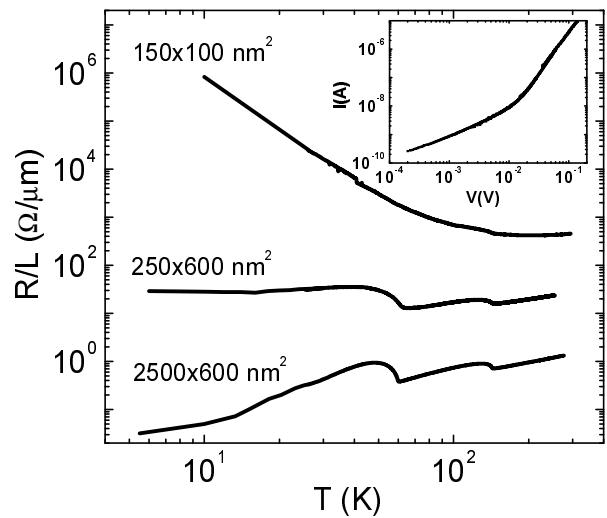}
\vskip -6cm
\caption{Temperature dependent four-probe resistance for a thin crystal of 
NbSe$_3$ (lower curve), a nanoconstriction (middle curve) and the two-terminal resistance
of a nanoconstriction in a different sample (top curve).
The estimated contact resistance is substracted for the top curve. The numbers 
give the transverse sizes (width $\times$ thickness).
The inset shows a low-temperature IV curve of a NbSe$_3$ nanoconstriction. 
$T=10$~K.} \label{thinRT}
\end{figure}

Reduction of the transverse sample sizes reveals interesting
finite-size effect
(Fig.~\ref{thinRT}). First of all, when the resistance per unit length,
$\rho=R(300{\rm~K})/L$, of thin cleaved (none-etched) crystals of NbSe$_3$ exceeds
0.1~k$\Omega/\mu$m, the resistance ratio, RRR, starts to decrease. When $\rho$
approaches to 1 k$\Omega/\mu$m, the low-temperature metallic conduction
($dR/dT<0$) converts into nonmetallic behavior ($dR/dT<0$). Further reduction of
the transverse sample sizes done by the FIB leads to further changes in the entire 
$R(T)$ dependence: disappearance of both Peierls transitions and appearance of a
nonmetallic behavior which can be approximated by the power law, $R\propto
T^{-\alpha}$, with $\alpha = 1$ - 3. Finally, the $R(T)$ curves of the
thinnest crystals have nothing in common with the initial $R(T)$ of thick 
samples.

The inset in Fig.~\ref{thinRT} shows a typical low-temperature I-V curve of a
nanoconstriction. It is nonlinear and the nonlinear part can be approximated
by the the power law, $I\propto V^\beta$ with $\beta \approx 3$.

\subsection{C/N/C junctions}

\begin{figure}
\vskip -1cm
\includegraphics[width=10cm]{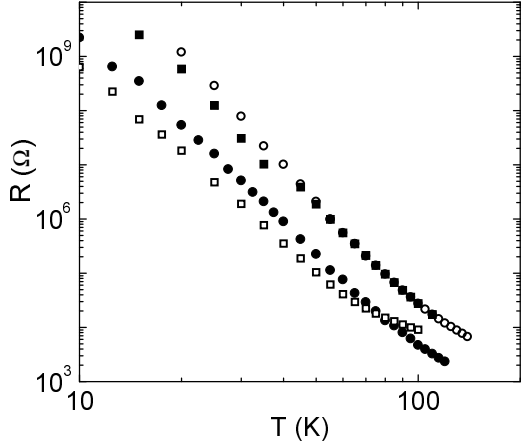}
\vskip -7cm
\caption{Double logarithmic plot of temperature variation of resistance of four
C/N/C junctions. Power laws with exponents between 5.5 and 6.5 are
observed over 5 decades in resistance.}
\label{loglogCNC} \end{figure}

Transport measurements of C/N/C junctions have been done in a two-terminal
configuration.  Before a junction is fabricated into a crystal, the $R(T)$ is
measured and demonstrates the expected behavior for NbSe$_3$.  After fabrication
of a metallic junction, the $R(T)$ measurement exhibits a steep rise in
resistance below 160~K, as illustrated in Fig.~\ref{loglogCNC}.  At low
temperatures the resistance increases dramatically and at $T\sim 10$~K it has
increased over 6 orders of magnitude when compared to the room temperature
resistance. Below 100~K, the logarithm of conductance is a linear function of
the logarithm of temperature (Fig.~\ref{loglogCNC}), i.e. $G(T)\propto
T^\alpha$.
Among 10 studied C/N/C junctions, three exponents lies in the region $2.9 \leq
\alpha \leq 3.4$, and seven belong to the interval $5.5 \leq \alpha \leq 6.5$.

\begin{figure}
\vskip -1cm
\includegraphics[width=10cm]{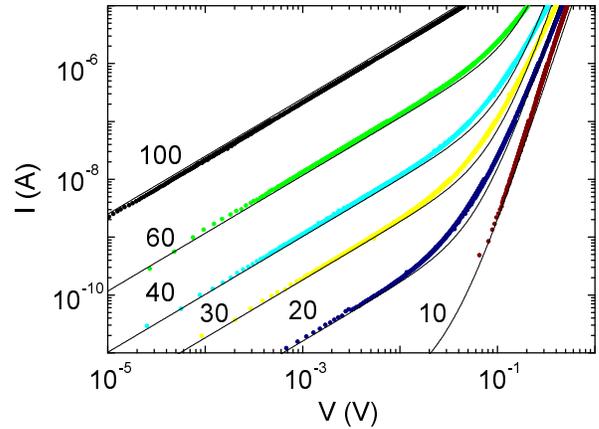}
\vskip -7cm
\caption{Typical temperature set of I-V curves of a sample with
C/N/C junction. Solid lines show the best fit by Eq.~\protect\ref{LLtheory}  with $C=5\times
10^{-20} $, $\gamma=0.108$, $\alpha=6$, $\beta=4$.}
\label{IVCNC}
\end{figure}

Fig.~\ref{IVCNC} shows a typical temperature set of current-voltage
characteristics (I-V curves) of a C/N/C junction. At temperatures above 100~K
the I-V curves are linear. Upon cooling, the I-V curves become more and more
nonlinear. Finally, at the lowest temperature studied, $T=10$~K, nonlinear
growth of the conduction, $G = I/V$, by four orders of magnitude is observed.

Four-terminal samples have been fabricated to investigate to what extend the
interface between NbSe$_3$ and platinum is important for the observation of behavior
described above. For this purpose two platinum probes are deposited on top of
the crystal surface by FIB deposition 40~$\mu$m apart of each other. At this stage
temperature variations of the sample resistance have been measured to verify 
its character before junction fabrication. Then a junction is fabricated on some
distance $L$ outside this 40 $\mu$m region. The platinum probes are used as 
voltage probes in a four-terminal setup. It has been found that for $L\geq
20$~$\mu$m
the junction fabrication does not affect the sample properties, whereas for
$L=5$ and 10~$\mu$m a rapid increase in resistance is observed when going down
in temperature. These measurements rule out the possibility that the
resistance growth is solely an interface phenomena. It was also found that both
etching alone and deposition of platinum on the top of the crystal alone do not
dramatically change the CDW properties of crystals as measured by electric
transport.  Thus, the resistance increase outside the junction region only occurs
when the crystal has been etched through and platinum has subsequently been
deposited.

\section{Discussion}

Our results show that quasi-one dimensional conductors with a CDW can
demonstrate the temperature and electric-field dependent conduction expected for
1D systems. The electric properties of nanoconstrictions in NbSe$_3$ reproduce
quantitatively the results obtained for long thin crystals of NbSe$_3$
\cite{ZZPM}. The electric properties of C/N/C junctions reveal qualitatively the
same type of behavior, but for much bigger transverse sizes of samples: 1D
behavior is observed even in samples with a width of the order of 10~$\mu$m.

A quasi-1D conductor can be considered as a set of metallic chains with LL
coupled by an interchain hopping $t_\perp$.  At $T<t_\perp$ the LLs are unstable
towards 3D coupling~\cite{Voit}, and the 3D ordered CDW may be formed.

Reduction of the transverse sizes of quasi-1D conductors decreases the stiffness
of the CDW.  This results in growth of dynamic CDW fluctuations.  In NbSe$_3$ a
growth of fluctuations is noticeable in thin samples with submicrometer
transverse sizes~\cite{Thorne}.  At a critical transverse size, when the
amplitude of fluctuation of a CDW phase difference between neighboring 
chains reaches a value of the order of $\pi$, the stiffness of the
shear deformation modes vanishes.  This leads to destruction of the 3D order of
the CDW, and the system starts to behave as a set of LLs, like it is expected in
a sliding LL phase at sufficiently small $t_\perp$~\cite{2DcoupledLL}.
An additional contribution may come from static CDW disorder.  Growth of static CDW
fluctuations may result from an increase of the surface contribution, as well as
from introduction of defects and impurities.  In this case one can also expect
frustration of the formation of a 3D-ordered CDW and recovery of LL properties.
It has been argued that the power laws expected for LLs, survive for the sliding
LL phase formed in 2D and 3D ordered chains of coupled LLs, at least in some
range of $t_\perp$~\cite{2DcoupledLL,SlidingLL}.

\begin{figure}
\vskip -2cm
\includegraphics[width=10cm]{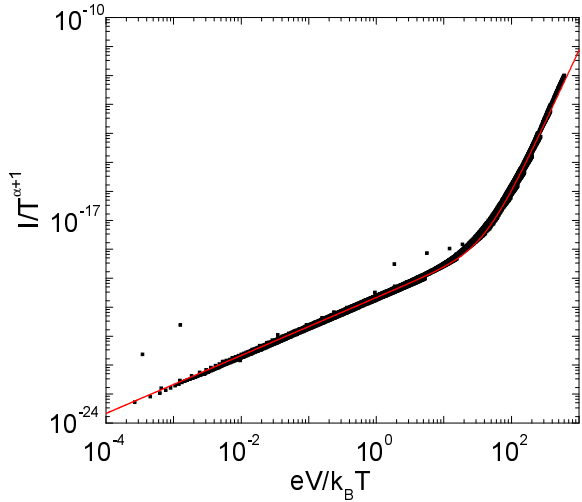}
\vskip -6cm
\caption{I-V curves  scaled in accordance with Eq.~\protect\ref{LLtheory}.
Solid lines show the best fit by Eq.~\protect\ref{LLtheory} with $C=5\times
10^{-20} $, $\gamma=0.108$, $\alpha=6$, $\beta=4$.}
\label{Master}
\end{figure}

If the scenario described above is correct, one could expect that the measured 
temperature
dependencies of the linear resistance, as well as I-V curves, can be fitted by
LL-like equations. According to Ref.~\cite{MasterCurve}, for a single-channel LL
the temperature set of I-V curves collapses into a master curve described by:
\begin{equation}
\frac{I}{ T^{\alpha+1}}=C {\rm sinh}\left(\gamma\frac{eV}{kT}\right)\left|\Gamma
\left(1+\frac{\beta}{2}+i\frac{\gamma}{\pi} \frac{eV}{k_BT}\right) \right|^2 ,
\label{LLtheory}
\end{equation}
where $\Gamma$ is the gamma-function, and $C$ and $\gamma$ are the fitting
parameters. One could expect that a similar equation, but with modified
$\alpha$ and $\beta$, describes stacks of LLs. Fig.~\ref{Master} shows the best
fit of the data of Fig.~\ref{IVCNC} by Eq.~\ref{LLtheory} with $C=5\times
10^{-20} $, $\gamma=0.108$, $\alpha=6$, $\beta=4$.  Apparently,
Eq.~\ref{LLtheory} describes the data well over the entire studied
temperature and voltage range.

The conduction of a set of coupled LLs is provided by both tunneling
along LLs through impurity barriers and tunneling between the chains with LLs.
The conduction exponents are $1/g-1$ for impurity barriers, and 
$(g+1/g-2)/2$ for interchain tunneling~\cite{TL_parameters}.  Both
conduction channels are connected in parallel.  The relative contribution of each conduction channel depends also on the strength of impurities, temperature and
the electric field. As the resistance along the chains grows faster upon
cooling than the resistance across the chains, interchain tunneling
dominates at low temperatures.

It is interesting to compare the observed exponents with the predictions of LL
theories.  The biggest observed exponent is $\approx 6$.  In the case of
intrachain tunneling we get 1/g=7.  Taking
$1/g\approx\sqrt{1+U/2E_F}$~\cite{LLReview,KF}, where $U$ denotes the potential
energy of electron-electron interaction and $E_F$ is the Fermi energy, one gets
$U\approx 100E_F$.  As $E_F\sim 1$~eV in NbSe$_3$~\cite{CDWReview}, the
respective $U\sim 10^2$~eV is too big (In the case of interchain tunneling, the
discrepancy is even more dramatic).  Similar discrepancies are observed if we
use a more accurate way of estimating $g$~\cite{TL_parameters}.  In accordance
with Ref.~\cite{TL_parameters}, to get $\alpha\approx 6$, we need $k_Fd\approx
0.1$, where $d$ is the effective wire diameter, and $k_F$ is the Fermi wave
vector.  As $k_F=Q_{CDW}/2=0.45$~\AA, where $Q_{CDW}$ is the CDW wave vector,
the diameter $d\approx 0.2$~\AA.  Such an effective wire diameter is too small.
Also for this estimation, an even more serious disagreement arises if we assume
interchain tunneling as the dominant mechanism of conduction.

The observed disagreement with predictions of LL theory for a single chain may
be caused by a renormalization of exponents by interchain
interactions~\cite{2DcoupledLL,SlidingLL}. We would like to note that the 
CDW-like component of the electron density~\cite{SabGind} ---which is expected 
to be large in our quasi-1D crystals --- can increase dielectrization.
Additional dielectrization (especially for C/N/C junctions) may come from
Friedel oscillations around impurity sites~\cite{FrOscComment,RomEgg}.  Further 
study is necessary to clarify the origin of the observed behavior.

\section{Conclusion}

Our results show that quasi-1D conductors with a CDW ground state can
demonstrate the temperature and electric-field dependent conduction expected for
1D systems.  For FIB-etched thin crystals we reproduced results obtained earlier
on thin un-etched crystals \cite{ZZPM}.  For C/N/C structures, the LL features are more pronounced:  we observed power laws up to 6 orders of magnitude in conduction. The
observed behavior corresponds to the destruction of 3D CDW ordering by
fluctuations and the recovery of the Luttinger liquid features of metallic chains.

\section{Acknowlegments}

We are grateful to Yu.A.~Firsov, V.A.~Sablikov and Yasha Gindikin for valuable
discussions, and to R.E.~Thorne for providing batches of high-quality crystals of NbSe$_3$.  This work is supported by the Dutch Foundation for Fundamental
Research on Matter (FOM), the Dutch Organization for Scientific Research (NWO),
the Russian Foundation for basic research (01-02-17771), and by MNTP ``Physics of Solid State Nanostructures'' 97-1052.  HSJvdZ is supported by the Dutch Royal Academy of Arts and Sciences (KNAW).

\end{document}